# CLINICAL CONCEPTS MIGHT BE INCLUDED IN HEALTH-RELATED MATHEMATIC MODELS

Dinamarca - Montecinos, José Luis
MD, PhD, Escuela de Medicina, Universidad de Valparaíso, Chile

Aguilar - Navarro, Sara
MD, PhD. Instituto Nacional de Ciencias Médicas y Nutrición Salvador Zubirán,
Ciudad de México, México

Runzer - Colmenares, Fernando
Universidad Científica del Sur, Lima, Perú

Morales - Eraso, Alexander
MSc, Universidad Cooperativa de Colombia, campus Pasto, Colombia

The construction of mathematical models that allow comprehensive approach of decision-making in situations of absence of robust evidence is important. While it is interesting to use models that are easy to understand, using values of direct interpretation, we analized a published index (COVID-19 Burden Index) and found it seems to be oversimplified. It is possible that the proposed index, with current data, could be useful in geographically and administratively narrowed places. But it is inaccurate to be applied throughout the process and in places as broad as American countries. It would be ideal to correct and refine the referred model, bearing in mind clinical concepts described, to take advantage of the proposal and generate a more accurate response, which can serve as an input both in the implementation of measures and in the prediction of the behavior of a pandemic like the current one. However, what we propose is to improve the accuracy of the model in terms of quantities and applicability, agreeing with the concept of "*stay at home*". The approach between complementary areas of knowledge should be the door that we must open to generate the new evidence we need. Mathematics should not dispense clinical sciences.

**Dear Editor:**

We have carefully read the article "Is a COVID19 Quarantine justified in Chile or USA right now?", published by RI González, F Muñoz, PS Moya and M Kiwi [1]. The construction of mathematical models that allow comprehensive approach of decision-making in situations of absence of robust evidence is important. As such, the approach from the exact sciences is matter of interest to the current pandemic.

Indeed, we have some examples of current models. In Mexico, an attempt has been made to predict the behavior of COVID-19 and estimate possible scenarios through Gauss model. With the Universidad de Guadalajara´s data, "contagion critical days" between March 20th-24th, were determined.

Assuming Y=1.0257x and R2=0.9846, and considering how the virus evolved in China and recent local data, the results using the equation, with 98% confidence, create some possible optimistic and critical scenarios. The model was built to estimate these scenarios at 5 days and not daily (as in Spain or Italy) given the fewest confirmed infected in Mexico [2].

While it is interesting to use models that are easy to understand, using values of direct interpretation, the proposed index ("COVID-19 Burden Index") seems to us to be oversimplified. This, because the authors minimize the impact of three relevant situations that are substantial issue of the integral process: First, in relation to a contextual justification, they use the assumption that different country-realities are comparable. Second, relative to the equation, they use a numerator greater than the actual one. Finally, in relation to the denominator, they do not incorporate fundamental clinical-epidemiological variables. All this leads to a possible overestimation of the total, with a risk of bias in the interpretation and applicability of the results. We will briefly develop these three elements in the same order in which we have just presented them.

Three realities (South Korea, United States of America and Chile) are put as comparable considering data related only to the COVID-19 incidence. It is not taken into account that South Korea is a small country [3] with just over 100,000 km² (109° according to land surface globally), with four cities (Seoul, Busan, Incheon and

Daegu) concentrating just over a third of the country's total population; United States of America [4,5] is a federal country with almost 10 million Km² (4°), in which the fifteen most populous cities do not account for even 10% of the country's total; while Chile [6] is a highly centralized country with approximately 750,000 Km² (38°) in which a single city (Santiago) concentrates one third of the country's total population **(Table N° 1)**.

**Table N° 1**
Comparison of population percentages and distribution between South Korea, Chile and the United States of America

| Country | Area in km² | Population | N Cities* |
|---|---|---|---|
| U.S.A. | 9.831.513 | 328.239.523[5] | 62 |
| CHILE | 756.945 | 18.751.405 | 1 |
| S. KOREA | 100.339 | 51.446.201 | 4 |

*N Cities: Cities that account for about a third of the total population

Analyzing and comparing the incidence of a disease, without considering the geographical and population realities of three very different territories constitutes a methodological limitation, too important not to be considered in the development of a model. Since the measures being implemented will be different according to geographical realities, the initial rate of contagion will be different, and the distribution of resources available to deal with the disease will be different. In the Mexican example, the total N of confirmed cases reached a value less than the daily N confirmed in Spain and Italy, in the same time elapsed since "case one". That is why we think it is correct to compare similar realities, especially in terms of population concentration according to territorial area, and according to the distribution of critical resources that exist in a given place and time.

Next, the numerator (15% of the confirmed population), equates to between 3 and 6.5 times the estimated number of people that will require mechanical ventilation [7]. There is no justification for why this high percentage was used. Had it initially been decided that the total N of confirmed cases were going to develop severe pneumonia? However, in most hospitals in Latin America -including Chile- this type of patient does not necessarily recharge intensive care beds, but specialized beds. This is a major factor to refine in the proposed model. Corrections that would bring the model closer to reality would be:

N confirmed – (N recovered + N deceased)
Or, in any case:
N confirmed × 0.05

In statistical terms, the problem is to use all of the N observations to obtain optimal estimates of the $\hat{\theta}_i$ of the $\theta_i$, considering the uncertainty introduced by the observation errors. This is a problem that Gauss describes as "the most important of the application of mathematics to natural philosophy" [2].

Finally, with regard to the denominator, the following elements should be well thought-out: The distribution of clinical resources that each country has to deal with a situation such as the current one, beyond the gross N. Relevant in all cases, is especially relevant in countries with strong organizational and administrative centralism, as Latin Americans: The distribution of resources is heterogeneous and does not have to relate to a particular N of contagions in the same place and time. On the other hand, there is a set of variables that makes a comparison between countries more complex. Beyond the effect that certain measures implemented may have generated on the shape of the infection, lethality and recoverability curves, there will always be population-specific variables that will end up affecting the development of these curves: genetic, nutritional and idiosyncratic elements. Variables of this type should be considered and quantified to be included in the denominator in the form of a corrective factor.

It is possible that the proposed index, with current data, could be useful in geographically and administratively narrowed places. But it is inaccurate to be applied throughout the process and in places as broad as American countries: Today, the reality in Chile shows that for several days we would be with a result "1" which, according to the model, should be interpreted as "collapse of the health system", which is still far from happening. It would be ideal to correct and refine the presented model, bearing in mind the concepts described, to take advantage of the proposal and generate a more accurate response, which can serve as an input both in the implementation of measures and in the prediction of the behavior of a pandemic like the current one. However, what we propose is to improve the accuracy of the model in terms of quantities and applicability, agreeing with the concept of "*stay at home*".

In conclusion, we believe that the approach between complementary areas of knowledge should be the door that we must open to generate the new evidence we need. Mathematics should not dispense clinical sciences, and the result of mutual synergy will always exceed the mere sum between the two. As long as evidence is built, we must avoid venturing conclusions that can promote or result in hasty decisions.

## References

1. González RI, Muñoz F, Moya PS, Kiwi M. Is a COVID19 Quarantine justified in Chile or USA right now? arXiv:2003.10879v1 [physics.med-ph] 24 Mar 2020.
2. Sprot D.A. (1978) Gauss's contributions to statistics. Historia Mathematica 5, p. 183-203.
3. Food and Agriculture Organization of the United Nations. Korea. http://www.fao.org/countryprofiles/index/en/?iso3=KOR
4. U.S. Census Bureau, Statistical Abstract of the United States: 2012, sección 6 Geography and Envi-ronment, table 358. Accesible en https://www.census.gov/library/publications/2011/compendia/statab/131ed.html
5. United States Census Bureau, County Population Totals: 2010-2019. Estimation 2019. Disponible en: https://www.census.gov/data/tables/time-series/demo/popest/2010s-counties-total.html#par_textimage_242301767
6. Oficina de Información Diplomática, Dirección General de Comunicación, Diplomacia Pública y Redes, Ministerio de Asuntos Exteriores, Unión Europea y Cooperación, Gobierno de España. Ficha país: República de Chile. Chile. Disponible en: http://www.exteriores.gob.es/Portal/es/SalaDePrensa/Paginas/FichasPais.aspx
7. Wei-jie Guan, Zheng-yi Ni, Yu Hu, Wen-hua Liang, Chun-quan Ou, Jian-xing He, Lei Liu, Hong Shan, Chun-liang Lei, David S.C. Hui, Bin Du, Lan-juan Li, et al. Clinical Characteristics of Coronavirus Disease 2019 in China. DOI: 10.1056/NEJMoa2002032.